\documentstyle[amsfonts,12pt]{article}
\renewcommand{\Re}{\mathop{\mathrm{Re}}}
\renewcommand{\Im}{\mathop{\mathrm{Im}}}
\renewcommand{\d}{{\mathrm {d}}}
\newcommand{\e}{{\mathrm {e}}}
\renewcommand{\i}{{\mathrm {i}}}
\begin{document}
\title{Schr\"odinger operators with complex potential but real spectrum}
\author{
Francesco Cannata\thanks{cannata@bo.infn.it}\\[2mm]
Dipartmento di Fisica and INFN, Via Irnerio 46, I-40126 Bologna, Italy\\[5mm]
Georg Junker\thanks{junker@theorie1.physik.uni-erlangen.de} 
and 
Johannes Trost\thanks{jtrost@theorie1.physik.uni-erlangen.de} \\[2mm]
Institut f\"ur Theoretische Physik, 
Universit\"at Erlangen-N\"urnberg,\\ 
Staudtstr.\ 7, D-91058 Erlangen, Germany\\[5mm] 
}
\maketitle
\begin{abstract}
Several aspects of complex-valued potentials generating
a real and positive spectrum are discussed. In particular, we construct
complex-valued potentials whose corresponding Schr\"odinger eigenvalue
problem can be solved analytically. \\[2mm]
\end{abstract}

\section{Introduction}
Quantum systems characterized by non-hermitian Hamiltonians are of interest in
several areas of theoretical physics. For example, in nuclear physics
\cite{BoMo69} one studies standard Schr\"odinger Hamiltonians with
complex-valued potentials, which in this connection are called optical
or average nuclear potentials. Non-hermitian interactions are also discussed
in field theories, for example, when studying Lee-Yang zeros
\cite{ItDr89}. Even in recent studies on localization-delocalization
transitions in superconductors \cite{FeZe97} and in the theoretical
description of defraction of atoms by standing light waves \cite{BeDe98}
non-hermitian Hamiltonians are of interest.

Recently, several authors \cite{Anetal97,BeBo97,BeBo98,BeMi98} have
studied standard one-dimensional Schr\"odinger Hamiltonians with 
complex-valued potentials giving rise to a real energy spectrum. Some of 
this work \cite{BeBo97,BeBo98} has been concentrated on the numerical study of
parity and time-reversal (PT) invariant Hamiltonians, and it is believed
\cite{Bendercomm}  that this invariance is a sufficient criterion for the
reality and positivity of the spectrum.
It is the aim of this note to construct via the Darboux method new complex
potentials for which the corresponding eigenvalue problem can be solved
exactly. These potentials are not necessarily PT invariant but still give rise
to a real and positive spectrum.

In order to make this note self-contained and also for the purpose of setting
up our notation we briefly review in the next section the basics of the
Darboux method. Using this method we then derive a rather general class of
complex potentials giving rise to the harmonic oscillator spectrum. In
particular, we note that this class of potentials also contains members which
are not PT invariant. In Section 4, we will discuss the Darboux approach for
the class of potentials $V(x)=-(\i x)^N/2$, which has recently been
discussed in some detail by Bender and Boettcher \cite{BeBo97,BeBo98}. 
Finally, in Section 5 we discuss the exactly solvable potential
$V(x)=\frac{1}{2}\exp\{2\i x\}$, which may be considered as a superposition of
the Bender-Boettcher potentials.


\section{The Darboux Method}
In this section we briefly review the method of Darboux \cite{Dar1882}, which
relates the spectral properties of a pair of standard Schr\"odinger
Hamiltonians 
\begin{equation}
  \label{1}
  H_{1/2}=-\frac{1}{2}\frac{\d^2}{\d x^2} + V_{1/2}(x)
\end{equation}
acting on the Hilbert space $L^2({\mathbb R})$.
Let us assume that for one of these Hamiltonians, say $H_1$, the spectral
properties are exactly known. That is, its eigenvalues $E_n$ and corresponding
eigenfunctions $\phi_n$ are known explicitly:
\begin{equation}
  \label{2}
  H_1 \phi_n(x)=E_n \phi_n(x)\;.
\end{equation}
For simplicity we will assume that $H_1$ has a purely discrete positive
spectrum enumerated by $n=0,1,2,\ldots$ such that $0<E_0 < E_1<E_2< \ldots $. 
If we now postulate that there exists a linear operator $A$ such that
\begin{equation}
  \label{3}
  A H_1 = H_2 A
\end{equation}
then the functions $\psi_n := A \phi_n \neq 0$ are obviously eigenfunctions
of the other Hamiltonian $H_2$, 
\begin{equation}
  \label{4}
   H_2 \psi_n(x)=E_n \psi_n(x),
\end{equation}
with the same eigenvalue $E_n > 0$. 

A rather general form for an intertwining operator obeying (\ref{3}) is given
by
\begin{equation}
  \label{5}
  A=\sum_{k=0}^{N}f_k(x)\,\frac{\d^k}{\d x^k}\;,
\end{equation}
where the $f_k:{\mathbb R}\to{\mathbb C}$, $k=0,1,\ldots, N-1$, 
are (at least twice)
differentiable functions to be determined via  the condition (\ref{3}) and
$f_N$ is an arbitrary constant.

The simplest non-trivial choice is $N=1$,\footnote{More general approaches for
  standard real-valued potentials can be found in \cite{others}.}
\begin{equation}
  \label{6}
  A=-\frac{\d}{\d x}+f(x)\;,\quad f:{\mathbb R}\to{\mathbb C}\;.
\end{equation}  
Putting this as an ansatz into eq.(\ref{3}) leads to ($f'=
\d f/\d x$ etc.) 
\begin{equation}
  \label{7}
  (V_2 - V_1 +f')\frac{\d}{\d x} - [(V_2 - V_1) f +
  V_1' - f''/2] {\bf 1} =0\;. 
\end{equation}
As $\frac{\d}{\d x}$ and the unit operator ${\bf 1}$
are linearly independent, 
their coefficients have to vanish simultaneously. That is, we have the two
conditions
\begin{equation}
  \label{8}
  V_2 = V_1 - f'\;,\qquad (V_2 -V_1)f+ V_1' - f''/2 =0\;.
\end{equation}
The first equation expresses the potential $V_2$ in terms of $V_1$ and
$f$, and thus allows to eliminate $V_2$ from the second one leading,
after an additional integration, to
\begin{equation}
  \label{9}
  f' +f^2 -2 V_1 = -2\varepsilon = const.
\end{equation}
Substitution $f=u'/u$ then results in the Schr\"odinger-like equation
\begin{equation}
  \label{10}
 \left( -\frac{1}{2}\frac{\d ^2}{\d x^2} + V_1(x) \right)  u(x)= \varepsilon
 u(x) 
\end{equation}
with in general complex integration constant $\varepsilon $. We will, however,
restrict ourselves to real $\varepsilon$ for reasons to be given below. Note
that we do not require square-integrability of $u$. We also note that in terms
of $f$ the two potentials read
\begin{equation}
\textstyle
  V_{1/2}(x)=\frac{1}{2}\,f^2(x)\pm \frac{1}{2}\,f'(x)+\varepsilon\;,
\label{10a}
\end{equation}
that is, for real $f$ the two Hamiltonians $H_{1/2}$ are SUSY partners
\cite{Ju96}. 

The above approach may be used to construct also complex-valued
potentials which generate a real-valued spectrum containing that of $H_1$.
To be more explicit, we start with a given real potential $V_1$, solve
eq.(\ref{10}) and thus obtain a new (in general complex) potential $V_2$,
which is given by 
\begin{equation}
  \label{11}
  V_2(x)=\left(\frac{u'(x)}{u(x)}\right)^2 - V_1(x) + 2 \varepsilon\;.
\end{equation}
Note that even for real $ \varepsilon$ the above potential may be complex by
choosing a complex linear combination of the two fundamental solutions of
(\ref{10}). This linear combination is not arbitrary because $u$ must not have
zeros on the real line in order to lead to a well-defined Hamiltonian $H_2$ on
$L^2({\mathbb R })$. Note that for $\varepsilon > E_0$  any regular
solution $u\in L^2({\mathbb R})$ will have zeros.
Hence, we will respect the condition $\varepsilon <E_0$, which is still not
sufficient for having a $u$ without zeros.

With the help of the intertwining relation (\ref{3}) it is now easy to see
that the strictly positive spectrum of $H_1$ forms a subset of the 
complete spectrum of $H_2$ with the corresponding eigenfunctions given by
\begin{equation}
  \label{11a}
  \psi_n(x)= C_n \,A\, \phi_n(x)= 
             C_n \left(-\phi_n'(x)+ \frac{u'(x)}{u(x)}\,\phi_n(x)
  \right)\;,
\end{equation}
where $C_n$ stands for a normalization constant defined by
$|C_n|^{-2}=\langle A\phi_n|A\phi_n\rangle$. 
Note that this normalization constant vanishes if
$\psi_n$ is not square integrable, implying that the corresponding eigenvalue
does not belong to the spectrum of $H_2$. This, however, happens only
for the square-integrable solutions of (\ref{10}), which we have eliminated
via the condition $\varepsilon  < E_0$. 

Noting that from (\ref{3}) follows $H_1(A^\dagger)^*=(A^\dagger)^*H_2$ where
$(A^\dagger)^*=\d/\d x +f(x)$, we realize that the above set of eigenfunctions
(\ref{11a}) is only complete on  
$L^2({\mathbb R})\backslash\mathrm{ker}\,(A^\dagger)^*$. Note that
$(A^\dagger)^*$ is a first-order differential operator and, therefore, the
dimension of 
$\mathrm{ker}\,(A^\dagger)^*$ may not exceed unity. In other words, $H_2$ may
have one additional eigenvalue which is below $E_0$. Note that 
$\mathrm{ker}\,(A^\dagger)^*$ is a one-dimensional subspace of
$L^2({\mathbb R})$ iff the 
differential equation  $(A^\dagger)^*\psi_\varepsilon=0$ has a solution in
this Hilbert space. This solution can explicitly be given
\begin{equation}
  \label{11b}
  \psi_\varepsilon (x)=\frac{C_\varepsilon}{u(x)}
\end{equation}
and it may easily be verified that it is an eigenfunction of
$H_2$ with eigenvalue $\varepsilon$, which belongs to the
spectrum of $H_2$ iff $\psi_\varepsilon\in L^2({\mathbb R})$. 

As conclusion of this section we note that the Darboux method may be
generalized such that one can construct complex potentials generating a real
spectrum  which is identical to that of a self-adjoint standard
Schr\"odinger Hamiltonian $H_1$, ${\mathrm spec}\, H_1=\{E_0,E_1,\ldots\}$. 
In some cases, via an appropriate choice of the
parameter $\varepsilon$ and the solution of (\ref{10}) this complex potential
may have an additional real eigenvalue $\varepsilon$ which lies below of the
spectrum of $H_1$. In the next section we will
demonstrate this approach for the case of the harmonic potential $V_1(x)=
\frac{1}{2}\,x^2$. 


\section{Complex potentials generating a harmonic\\ spectrum}
In this section we will consider as an explicit example the harmonic oscillator
potential $V_1(x)= (1/2)x^2$. The spectral properties of the
corresponding Schr\"odinger Hamiltonian $H_1$ are well known:
\begin{equation}
  \label{12}\textstyle
  E_n= n + \frac{1}{2}\,,\quad 
  \phi_n(x)=[\sqrt{\pi}\,2^n n!]^{-1/2}H_n(x)\exp\{-x^2/2\}\;,
\end{equation}
where $H_n$ denotes the Hermite polynomial \cite{MOS66} of degree
$n\in\{0,1,2,\ldots\}$. 
The most general solution of the Schr\"odinger-like equation (\ref{10}) can be
given in terms of confluent hypergeometric functions \cite{MOS66}
\begin{equation}
  \label{13} \textstyle
   u(x)= \e^{-x^2/2}\left[
    \alpha\;_{1}F_{1}(\frac{1-2\varepsilon}{4},\frac{1}{2},x^2) +
    \beta x\;_{1}F_{1}(\frac{3-2\varepsilon}{4},\frac{3}{2},x^2)\right]\;.
\end{equation}

As $u$ should not have any zero on the real line $\alpha$ must not vanish and,
therefore, can be set equal to unity without loss of generality. 
Note that an overall factor in $u$ is irrelevant for the relevant 
formulas (\ref{11})-(\ref{11b}). 
From the general discussion of the last section we also have the 
condition $\varepsilon < E_0=1/2$. Finally, we have to determine for which
values of the remaining parameter $\beta\in{\mathbb C}$ the general solution
(\ref{13}) has no zeros on the real line. These conditions
can be extracted \cite{JuRo97,JuRo98} from the asymptotic behaviour
\begin{equation}
  \label{15}
  u(x)=\frac{\exp\{x^2/2\}}{|x|^{1/2+\varepsilon}}
\left[\frac{\Gamma(\frac{1}{2})}{\Gamma(\frac{1-2\varepsilon}{4})}
+\beta\, \frac{\Gamma(\frac{3}{2})}{\Gamma(\frac{3-2\varepsilon}{4})}\,
\frac{x}{|x|} + O(1/|x|)\right]\;.
\end{equation}
For $\beta\in{\mathbb R}$ a strictly positive solution is only possible if
\begin{equation}
  \label{15a}
|\beta|<\beta_c(\varepsilon):= 2\,
\frac{\Gamma(\frac{3}{4}-\frac{\varepsilon}{2})}
     {\Gamma(\frac{1}{4}-\frac{\varepsilon}{2})}\;.
\end{equation}
Violation of this condition results in singularities in $V_2$. See ref.\
\cite{JuRo97,JuRo98} for details. 
However, for $\beta\in{\mathbb C}/{\mathbb R}$ the solution $u$ does not have
any zero. That is, the allowed values for $\beta$ are given by the complex
plane with 
the exception of the two cuts $]-\infty,-\beta_c(\varepsilon)]$ and
$[\beta_c(\varepsilon), \infty[$ on the negative and positive real line,
respectively. In addition, from (\ref{15}) we can read off that 
$\psi_\varepsilon =C_\varepsilon /u$ is square
integrable in the allowed ranges of the parameters. In other words, we have an
additional eigenvalue $\varepsilon< 1/2$ in $H_2$. In Figure 1 we show the
real part (a) as well as the 
imaginary part (b) of the potential $V_2$ for $-3<\varepsilon<1/2$ and
$\beta=\i$. We note here that for all $\beta\in
\i {\mathbb R}$ the potential is 
invariant under  PT transformations \cite{BeBo97}, that is, $V_2^*(-x)=V_2(x)$
and, therefore, also the Hamiltonian $H_2$ has this symmetry. However, 
other complex values of $\beta$, which are not purely imaginary, are also
admissible 
and thus provide examples of non-PT invariant potentials generating the real
harmonic oscillator spectrum with the additional ground-state energy
$\varepsilon$. 

In Figure 2 and 3 we give several examples for fixed $\varepsilon=-1/2$ and
typical values of  $\beta$. Note that for this particular value of
$\varepsilon$ the solution (\ref{13}) can be expressed in terms of the error
function \cite{MOS66}
\begin{equation}
\label{15aa}
 u(x)=\e^{x^2/2}\left[1+\beta\,\frac{\sqrt{\pi}}{2}\,{\mathrm
 Erf}(x)\right]\;,\quad 
 \beta_c(-1/2)=2/\sqrt{\pi}\approx 1.128\;.
\end{equation}
This example explicitly shows that for all
\begin{equation}
  \label{15b}\textstyle
  \beta\in{\mathbb C}\backslash \left\{]-\infty,-\frac{2}{\sqrt{\pi}}]\cup
  [\frac{2}{\sqrt{\pi}},\infty[\right\}
\end{equation}
the solution given in (\ref{15aa}) does not have zeros on the real axis,
because its imaginary part vanishes only at $x=0$ 
where its real part is obviously non-zero. Figure 2 shows the real (a) as well
as the imaginary (b) part of $V_2$ for $\Re\,\beta=0.5$ and
$\Im\,\beta\in[-2,2]$, thus crossing the real axis of the complex
$\beta$-plane in the allowed region. 
Figure 3 shows the same for $\Re\,\beta=2$ and 
$\Im\,\beta\in[-1,0]$, that is, reaching the cut of the $\beta$-plane,
which is lying on the positive real axis, from below. Here a singular
behaviour is clearly visible, when $\beta$ approaches the cut.

We conclude this section in summarizing the spectral properties of $H_2$:
\begin{equation}
  \label{16}
  \begin{array}{l}
    {\mathrm spec}\,H_2 = \{\varepsilon,
    \frac{1}{2},\frac{3}{2},\frac{5}{2},\ldots\}\;,\quad 
    \psi_\varepsilon(x)=C_\varepsilon/u(x)\;,\\[2mm]
 \displaystyle
 \psi_n(x)=\frac{\exp\{-x^2/2\}}{[\sqrt{\pi}\,2^{n+1} n!
 (n+1/2-\varepsilon)]^{1/2}}
 \left[H_{n+1}(x)+\left(\frac{u'(x)}{u(x)}-x\right)H_n(x)\right]\;,
  \end{array}
\end{equation}
where the normalization constants have been determined via\footnote{Note that
  this is a priori only true for $\beta\in{\mathbb R}$. But as the result is
  independent of $\beta$ we may analytically continue also to complex values.}
$|C_n|^{-2}=\langle A\phi_n|A\phi_n\rangle
=2\langle\phi_n|H_1-\varepsilon|\phi_n\rangle=2(E_n-\varepsilon)$.
We note again that the spectrum
of $H_2$ is real and bounded below for any finite value of $\varepsilon<1/2$.
This remains true even in the case of complex $\beta$ where $H_2$ is neither
self-adjoint nor PT invariant in general. Because of $\lim_{|x|\to\infty}
\Im\,V_2(x)=0$, the eigenvalue problem of $H_2$ is well-defined on
$L^2({\mathbb R})$. Let us also remark that following the approach of
\cite{JuRo97} we are able to define ladder operators for the non-hermitian
Hamiltonian $H_2$ closing a quadratic algebra.

\section{Application to the Bender-Boettcher potentials}
The Darboux method reviewed in Section 2 is also applicable to complex
potentials. 
Here we will consider the class of potentials,\footnote{Note that the special
  case $N=2$ corresponds to the harmonic oscillator discussed in the previous
  section.} 
$N\geq 2$, 
\begin{equation}
  \label{21}
  V_1(x):=-\frac{1}{2}\,(\i x)^N\;,
\end{equation}
which has recently been discussed extensively by Bender and Boettcher
\cite{BeBo97,BeBo98}. These authors have shown (numerically) that for $N\geq
2$ the above 
potential generates a real and strictly positive spectrum on an appropriate
Hilbert space, which is taken to be the linear vector space of square
integrable functions on some contour in the lower complex
$x$-plane. This contour has to lie in some sectors centered about anti-Stokes
lines and bounded by Stokes lines for $\Re\,x\to\pm\infty$
\cite{BeBo97}. For simplicity, we will assume that this contour, which is the
one-dimensional configuration space of the problem, approaches the anti-Stokes
lines, that is, 
\begin{equation}
  \label{22}
  \lim_{\Re\,x\to\pm\infty}\frac{x}{|x|}=
  \exp\left\{-\i \frac{\pi}{2}\pm \i  \frac{2\pi}{N+2}\right\}\;.
\end{equation}

For the present case the Schr\"odinger-like equation (\ref{10}) cannot be
solved analytically for 
arbitrary $\varepsilon$. However, for the special case $\varepsilon=0$, which
is below the ground-state energy of $H_1$, this
equation reduces to Bessel's differential equation. In other words, for this
special case we can express the solution of (\ref{10}) in terms of modified
Bessel functions \cite{MOS66}
\begin{equation}
  \label{23}
  u(x)=z^\nu\left[\alpha\, I_{\nu}(z)+\beta\, K_{\nu}(z)\right]\;,
\end{equation}
where we have set
\begin{equation}
  \label{24}
  z:=\frac{2}{N+2}\,(\i x)^{(N+2)/2}\;,\quad \nu:=\frac{1}{N+2}\;.
\end{equation}
Note that the anti-Stokes lines in the $x$-plane are mapped onto the negative
real line in the $z$-plane (this is the cut of the Bessel functions) such that
\begin{equation}
  \label{25}
  \lim_{\Re\, x \to \pm\infty}\frac{z}{|z|} = \e^{\pm \i  \pi}\;.
\end{equation}
The new potential generated via the Darboux method now reads
\begin{equation}
  \label{26}
  V_2(x)=-\frac{1}{2}\,(\i x)^N
  \left[2\left(
  \frac{\alpha\, I_{\nu-1}(z)-\beta\, K_{\nu-1}(z)}
       {\alpha\, I_{\nu}(z)+\beta\, K_{\nu}(z)}\right)^2-1\right]
\end{equation}
and approaches for $|x|\to \infty$ asymptotically the original potential $V_1$
given in (\ref{21}), that is, $V_2(x)=V_1(x) [1+O(1/|x|)]$. As a consequence,
the new potential $V_2$ has the same 
Stokes and anti-Stokes lines as the original one and, therefore, both
Hamiltonians can be defined on the same Hilbert space. In other words, the
Darboux transformation leaves the Hilbert space invariant.

We remark that
for appropriate choices of the parameters $\alpha$ and $\beta$ the Hamiltonian
$H_2$ has an additional eigenvalue $\varepsilon=0$, which is its ground-state
energy. This, in fact, corresponds to a situation where SUSY is unbroken. Note
that the Hamiltonians $H_{1/2}$ are SUSY partners \cite{Ju96}. In the case of
broken SUSY $H_2$ has the same spectrum as $H_1$.
By allowing for complex $\alpha$ and $\beta$ it is also possible to
construct potentials $V_2$ which are not PT-invariant. 
Nevertheless, the spectrum is identically with that of $V_1$ which, for $N\geq
2$ is strictly positive and discrete \cite{BeBo97}.
Note that in the
present context PT-invariance of a potential $V$ means $V(x)=V^*(-x)$ for all
$x\in{\mathbb R}$.  


\section{Another exactly solvable complex potential}
In this section we will consider the eigenvalue problem
associated with the complex potential $V(x)=\frac{1}{2}\,\exp\{2\i x\}$, which
may be considered as a superposition of all Bender-Boettcher potentials. 
This potential is periodic, $V(x+\pi)=V(x)$, and, in some sense, simulates a
proper regularized large $N$ limit of (\ref{21}).  
The corresponding eigenvalue problem can be write in the form  
\begin{equation} 
\psi''(x) + \left(2 \, E - \e^{2\i x}\right) \psi(x) = 0\;,
\label{eq:schroe1} 
\end{equation} 
which in turn is reducible to Bessel's differential equation \cite{MOS66}.
Pairwise linearly independent solutions are, for example, given by
\begin{eqnarray} 
&&\psi_{1}(x) = H_{\sqrt{2E}}^{(1)}\left(\e^{\i x}\right)\;, \quad 
\psi_{2}(x) = H_{\sqrt{2E}}^{(2)}\left(\e^{\i x}\right) \; , 
\nonumber \\   
&&\psi_{3}(x) = J_{\sqrt{2E}}\left(\e^{\i x}\right)\;, \quad 
\psi_{4}(x) = \left\{
\begin{array}{lll}
Y_{\sqrt{2E}}\left(\e^{\i x}\right) \ \mbox{for $E\geq 0$} \\[1mm]
J_{-\sqrt{2E}}\left(\e^{\i x}\right) \ \mbox{for $E<0$} 
\end{array}
\right. .
\label{eq:solpsi} 
\end{eqnarray} 

In the complex upper-half $x$-plane, that is $\Im x>0$, the
normalizable solution for $E>0$ is given by $\psi_{3}$. For $E<0$  
we have $\psi_{3}$ and $\psi_{4}$ as normalizable solutions in
the sectors $0<{\mathrm arg}\,x<\pi/2$ and $\pi/2<{\mathrm arg}\,x<\pi$,
respectively. 
There are no further restrictions on the parameter $E$.
Hence, the spectrum of the corresponding Hamiltonian is given by the 
complete real line and thus is not bounded from below.
In other words, in the complex upper-half $x$-plane the potential under 
consideration does not model a physically relevant system.
The same holds for $x\in{\mathbb R}$, where no normalizable solutions can 
be found. 

In the complex lower-half $x$-plane, $\Im x<0$, the argument of the Bessel 
functions $z:= \e^{\i x}=\e^{|\Im x|}\e^{\i\Re x}$
becomes infinitely large for $\Im x\to-\infty $. In order to find the
physically acceptable  
sectors in the lower-half plane we consider the asymptotic behaviour of the 
pair $\psi_{1}$ and $\psi_{2}$,
\begin{eqnarray} 
\psi_{1}(x)  &=&  \sqrt{\frac{2}{\pi}}\, \e^{-\i x/2}  
\exp\left[\i\left(\e^{|\Im x|}\e^{\i\Re x}
-\frac{\pi}{2}\sqrt{2E}  -\frac{\pi}{4}\right)\right] 
(1+O(1/z))\;, \nonumber \\  
\psi_{2}(x)  &=&  \sqrt{\frac{2}{\pi}}\, \e^{-\i x/2}  
\exp\left[-\i \left(\e^{|\Im x|} \,\e^{\i\Re x}
-\frac{\pi}{2}\sqrt{2E} - \frac{\pi}{4}\right)\right] (1+O(1/z))\; . 
\label{eq:solpsiasymH} 
\end{eqnarray} 
The second exponential in (\ref{eq:solpsiasymH}) is dominating the 
asymptotic behaviour of $\psi_{1}$ and $\psi_{2}$. Which one of these two 
will be the exponentially decreasing solution depends on the sign of the 
imaginary part of $\e^{\i\Re x}$. As a consequence, the complex lower-half 
$x$-plane is divided into vertical stripes defined by
\begin{equation} 
S_{l}=\{x\in{\mathbb C}|\Im x<0, l\pi\leq\Re x<(l+1)\pi]\;,\quad l\in{\mathbb 
Z}\;. 
\end{equation} 
For $x\in S_{2l}$ we find that $\psi_{1}$ is the normalizable solution.
Whereas for $x\in S_{2l+1}$ the normalizable solution is given by $\psi_{2}$.
At the borders between neighbouring stripes, that is, for $\Re x =l\pi$ the 
first exponential in both relations (\ref{eq:solpsiasymH}) guarantees square 
integrability of both solutions. 

The above discussion shows that the contour along which we have to define our 
quantum system has to be confined to one of these strips for asymptotically
large $x$. In  
other words, the anti-Stokes lines for the present potential lie vertical in 
the lower-half plane and are equally spaced by a distance $\pi$. In fact, this 
is what one expects in the limit $N\to\infty $ in (\ref{22}). Hence, there are 
infinitely many possible choices in defining a proper quantum eigenvalue 
problem for the present potential. In what follows we will discuss some 
typical cases.

Without loss of generality let us assume that the contour starts at $\Im x
=-\infty$ , say, in $S_0$. Hence, $\psi_{1}$ is the proper solution. If we now
choose our configuration space such that this contour also ends in the same
sector we will not attain any restrictions on $E$. That is, we have again the
unphysical situation of a quantum system with a spectrum unbound from below.
Therefore, in order to have a proper quantum system we must demand that the
contour ends in some other sector, say, $S_m$ with $m>0$.
In this case we have to investigate the analytic continuation of $\psi_1$ into
this sector, which is given by \cite{MOS66} 
\begin{equation} 
\sin(\nu\pi) \, H_{\nu}^{(1)}(z\, \e^{m\pi \i }) = 
-\sin\left[(m-1)\nu\pi\right] \, H_{\nu}^{(1)}(z) - \e^{-\nu\pi \i } \, 
\sin(m\nu\pi) \, H_{\nu}^{(2)}(z)\;. 
\label{eq:anacon} 
\end{equation} 
where $E=\nu^2/2$.
For $m=2l>0$ we have to avoid an admixture of $H^{(2)}$, which is the 
exponentially growing solution in $S_{0}$ and $S_{2l}$. Hence we must demand
that $\sin(2l\pi\nu)=0$ and $\sin(\pi\nu) \neq 0 $. In the particular case
$l=1$ this results in the condition $\nu=n+\frac{1}{2}$, $n=0,1,2,3,\dots$,
giving rise to the positive discrete spectrum  
\begin{equation} 
E_n = \frac{1}{2} \, \left(n+\frac{1}{2}\right)^{2} \;,\quad n\in{\mathbb
  N}_0\;. 
\label{eq:quant} 
\end{equation} 
In the general case $m=2l$ we arrive at the conditions $2l\nu\in{\mathbb
N}$ and $\nu\notin{\mathbb N}$. For example, for $l=2$ we have 
$\nu=\frac{1}{4},\frac{1}{2},\frac{3}{4},\frac{5}{4},\frac{3}{2},\dots$,
leading to a ``perforated'' spectrum
\begin{equation}
  E_n=\frac{1}{8l^2}(n+1)^2\;,\quad n\in{\mathbb N}_0\backslash
\{2l-1, 4l-1,\ldots\}\;.
\label{eq:quantgen}
\end{equation}
 
Now we investigate the case $m=2l+1$ for which $H_{\nu}^{(2)}$ represents the
normalizable solution in the corresponding sector. For the special case $l=0$,
that is, the 
contour ends in the neighbouring sector $S_1$ we have to invoke the connection
formula \cite{MOS66}
\begin{equation} 
H^{(1)}_{\nu}(z\,\e^{\pi \i }) = - \e^{-\nu\pi \i}H^{(2)}_{\nu}(z)\;,
\label{eq:continuation}
\end{equation} 
which does not yield any condition on $\nu$. In other words, the corresponding
spectrum is unbound from below and thus unphysical. For $l\geq 1$ we again
make use of (\ref{eq:anacon}) leading to $2l\nu\in{\mathbb N}$ with 
$\nu\notin{\mathbb N}$. These conditions are identically to those
already found in the case $m=2l$. Hence, we will obtain the same spectrum
(\ref{eq:quantgen}).

The remaining part of this section is devoted to a short semiclassical
analysis of the problem (\ref{eq:schroe1}). 
If we look at the potential at the left border of $S_{0}$, that is 
$\Re x = 0$, the potential has the form $V(-\i R)=\frac{1}{2}\e^{2R}$
and thus exhibits a classical turning point for positive $E$. 
At the left border of $S_{m}$, i.e.\ $\Re x = m\pi$, the potential looks the
same because of its periodicity.   
For a semiclassical analysis we need two complex classical turning
points given, for example, by $x_{1} = -\i \ln k$ and $x_{2} = m\pi -\i \ln
k$, $m\geq 1$. They are the solutions of $k^{2}-\e^{2\i x} = 0$ with
$k^{2}=2E$. The semiclassical quantization condition is conventionally given by
\begin{equation} 
I=\int_{x_{1}}^{x_{2}} \d x \sqrt{k^{2}-\e^{2\i x}}=
\pi\left(n+\frac{1}{2}\right) \;,\quad n\in{\mathbb N}_0\;.
\label{eq:action}
\end{equation} 
In order to get a real value for the integral we have to integrate along the 
horizontal line between $x_{1}$ and $x_{2}$ in the complex plane
\cite{BeBo97}. Taking into account the periodicity of the integrant and
changing the integration variable to $z=\e^{2\i x}/k^2$  
the integral can easily be calculated and yields $I = m \pi k$. 
Therefore the semiclassical approximation leads to
\begin{equation} 
E^{\rm{sc}}_n= \frac{1}{2m^2} \, \left(n+\frac{1}{2}\right)^{2} \;,\quad
n\in{\mathbb N}_0\;,
\end{equation} 
which, in fact, is very similar to the exact result (\ref{eq:quantgen}) 
valid for both cases $m=2l$ and $m=2l+1$. 
However, we should note two essential failures of the semiclassical
approximation. Firstly, it gives rise to a discrete spectrum even in the
case $m=1$ where the exact treatment has led to an unbound spectrum. Secondly,
the semiclassical approach also fails to reproduce the perforation in the
spectrum found for the cases $m\geq 2$, c.f.\  (\ref{eq:quantgen}).

\section*{Acknowledgement}
One of us (G.J.) would like to thank Carl Bender for his valuable comments on
an earlier version of this manuscript.

\vfill
\centerline{\Large\bf Figure Captions}
\begin{itemize}
\item[Fig.\ 1:] Real (a) and imaginary (b) part of potential (\ref{11})
  with (\ref{13}) for various
  values of $\varepsilon$ and fixed $\beta=\i$. For purely imaginary $\beta$
  this potential is PT invariant.
\item[Fig.\ 2:] Same as Figure 1 but for fixed $\varepsilon=-1/2$ and various
  complex values of $\beta$ with $\Re\,\beta =0.5$. Note that for these
  values of parameters the potential is not PT invariant.
\item[Fig.\ 3:] Same as Figure 2 with $\Re\,\beta =2$. A singular
  behaviour emerges as $\beta$ approaches the real axis.

\end{itemize}


\begin{thebibliography}{99}
\bibitem{BoMo69} A.\ Bohr and B.R.\ Mottelson, Nuclear Structure, Vol.\ I,
  Sect.\ 2.4, (W.A.\ Benjamin Inc., New York, 1969).
\bibitem{ItDr89}C.\ Itzykson and J.-M.\ Drouffe, Statistical field theory,
  Vol.\ 1, Sect.\ 3.2.3, (Cambridge University Press, Cambridge, 1989).
\bibitem{FeZe97} J.\  Feinberg and  A.\ Zee, cond-mat/9706218.
\bibitem{BeDe98} M.V.\ Berry and D.H.J.\ O'Dell, J.\ Phys.\ A {\bf 31} (1998)
  2093. 
\bibitem{Anetal97}A.A.\ Andrianov, F.\ Cannata, J.-P.\ Dedonder and M.V.\
  Ioffe,   {\it SUSY quantum mechanics with complex superpotentials and real
  spectra}, preprint (1997).
\bibitem{BeBo97}C.M.\ Bender and S.\ Boettcher, physics/9712001 to appear in
  Phys.\ Rev.\ Lett.
\bibitem{BeBo98}C.M.\ Bender and S.\ Boettcher, J.\ Phys.\ A {\bf 31} (1998)
  L273.  
\bibitem{BeMi98}C.M.\ Bender and K.A.\ Milton, physics/9802184.
\bibitem{Bendercomm} C.M.\ Bender, private communication.
\bibitem{Dar1882}G.\ Darboux, Comptes Rendus Acad.\ Sci.\ (Paris) {\bf 94}
  (1882) 1456. 
\bibitem{others}A.A.\ Andrianov, M.V.\ Ioffe, F.\ Cannata and J.P.\ Dedonder,
  Int.\ J.\ Mod.\ Phys.\ A {\bf 10} (1995) 2683.\\ 
V.G.\ Bagrov and B.F.\ Samsonov, Teor.\ Mat.\ Fiz.\ {\bf 104}
  (1995) 356.\\
 V.G.\ Bagrov and B.F.\ Samsonov, J.\ Phys.\ A {\bf 29} (1996) 1011.\\
D.J.\ Fern\'{a}ndez C., M.L.\ Glasser and L.M.\ Nieto, Phys.\ Lett.\ A {\bf
  240} (1998) 15.
\bibitem{Ju96}G.\ Junker, Supersymmetric Methods in Quantum and Statistical
  Physics, (Springer-Verlag, Berlin, 1996).
\bibitem{MOS66}W.\ Magnus, F.\ Oberhettinger and R.P.\ Soni, Formulas and
  Theorems for the Special Functions of Mathematical Physics, 3rd ed.,
  (Springer-Verlag, Berlin 1966).
\bibitem{JuRo97}G.\ Junker and P.\ Roy, quant-ph/9709021 to appear in Yad.\
  Fiz. 
\bibitem{JuRo98}G.\ Junker and P.\ Roy, quant-ph/9803024.


\end{thebibliography}
\end{document}